\documentclass[twocolumn,prd]{revtex4}%
\usepackage{graphicx}
\usepackage{amsmath}
\usepackage{amsfonts}
\usepackage{amssymb}%
\providecommand{\U}[1]{\protect\rule{.1in}{.1in}}
\newcommand{\f}{\begin{equation}}
\newcommand{\ff}{\end{equation}}
\newcommand{\fa}{\begin{eqnarray}}
\newcommand{\ffa}{\end{eqnarray}}

\begin{document}
\title{Thermodynamics of modified black holes from gravity's rainbow}
\author{Yi Ling$^{1,2}$}\email{yling@ncu.edu.cn}
\author{Xiang Li$^{1,2}$}\email{xiang.lee@163.com}
\author{Hongbao Zhang$^{3,1}$}\email{hbzhang@pkuaa.edu.cn}
\affiliation{${}^1$ Center for Gravity and Relativistic
Astrophysics, Department of Physics, Nanchang University, 330047,
China}
\affiliation{%
${}^2$ CCAST (World Laboratory), P.O. Box 8730, Beijing
   100080, China}
\affiliation{ ${}^3$ Department of Physics, Beijing Normal University, Beijing, 100875, China\\
    Department of Astronomy, Beijing Normal University, Beijing, 100875, China}

\begin{abstract}
We study the thermodynamics of modified black holes proposed in
the context of gravity's rainbow. A notion of intrinsic
temperature and entropy for these black holes is introduced. In
particular for a specific class of modified Schwarzschild
solutions, their temperature and entropy are obtained and compared
with those previously obtained from modified dispersion relations
in deformed special relativity. It turns out that the results of
these two different strategies coincide, and this may be viewed as
a support for the proposal of deformed equivalence principle.

\end{abstract}

\pacs{03.30.+p, 04.60.-m, 04.70.Dy, 04.60.Pp} \maketitle

\section{Introduction}

Doubly special relativity is a deformed formalism of special
relativity to preserve the relativity of inertial frames while at
the same time keep Planck energy as an invariant scale, namely a
universal constant for all inertial
observers\cite{Amelino00ge,Amelino00mn,Amelino03ex,Amelino03uc,Magueijo01cr,Magueijo02am}.
This can be accomplished by a non-linear Lorentz transformation in
momentum space, which leads to a deformed Lorentz symmetry such
that the usual energy-momentum relations or dispersion relations
in special relativity may be modified with corrections in the
order of Planck length. Modified dispersion relations (MDR) can
also be derived in the study of semi-classical limit of loop
quantum
gravity\cite{Gambin98it,Smolin02sz,Sahlmann02qk,Smolin05cz}.
Experimentally such modifications may be responsible for threshold
anomalies of ultra high energy cosmic rays and Tev
photons\cite{Colladay98fq,Coleman98ti,Amelino00zs,Jacobson01tu,Myers03fd,Jacobson03bn}.

Recently in\cite{Magueijo02xx} Magueijo and Smolin argued that
this formalism may be generalized to incorporate the curvature of
spacetime. They proposed a deformed equivalence principle of
general relativity, stating that the free falling observers who
make measurements with energy E will observe the same laws of
physics as in modified special relativity. One important
implication of this idea is that there is no single classical
geometry of spacetime probed by a particle moving in it when the
effects of the probe itself are taken into account. On the
contrary, the spacetime is described by a one parameter family of
metrics which may depend on the energy of this particle, forming a
``rainbow'' metric. Consequently the connection and curvature are
energy dependent such that the usual Einstein's equations is
replaced by a one parameter family of equations. As specific
examples the modified version of FRW solution and Schwarzschild
solution to these equations have been presented in
\cite{Magueijo02xx} as well.

In this paper we intend to study the thermodynamics of modified
black hole solutions, aiming to test the idea of deformed
equivalence principle proposed above. As in \cite{LHL} we have
investigated the impacts of modified dispersion relations on black
holes. we find that the temperature as well as the entropy of
black holes receive corrections due to the modification of
energy-momentum relations of photons. Then the question is whether
these results are consistent with the proposal of % results in the
%context of
gravity's rainbow. We present an affirmative answer to this
question in this paper. Starting from the modified Schwarzschild
black hole solutions, we firstly obtain an energy dependent
temperature through the calculation of surface gravity, then we
propose the notion of intrinsic temperature as well as intrinsic
entropy for these black holes.  Comparing these results with the
ones obtained in \cite{LHL}, we argue that these quantities
obtained through gravity's rainbow and modified dispersion
relations are exactly the same and this coincidence may be viewed
as a support for the proposal of deformed equivalence principle.

\section{Rainbow metric and modified black hole solutions}

In this section we briefly review the rainbow metric proposed in
\cite{Magueijo02xx}. In the context of deformed or doubly special
relativity\cite{Magueijo02am}, the invariant of energy and
momentum in general may be modified as \f
E^2f_1^2(E,\eta)-p^2f_2^2(E,\eta)=m_0^2,\ff where $f_1$ and $f_2$
are two functions of energy from which a specific formulation of
boost generator can be constructed and $\eta$ is a dimensionless
parameter. The correspondence principle requires that $f_1$ and
$f_2$ approach to unit as $E/M_p\ll 1$. Since these theories are
typically formulated in momentum space and the transformation laws
are no longer linear, the definition of a dual space or position
space is non-trivial. One possible strategy is to require that the
contraction between momenta and infinitesimal displacement be a
linear invariant. \f dx^{\mu}p_{\mu}=dtE+dx^ip_i.\ff As a result,
the dual space is endowed with an energy dependent invariant which
is called a rainbow metric \f ds^2=-{1\over
f_1^2(E,\eta)}dt^2+{1\over f_2^2(E,\eta)}dx^2.\ff The rainbow
metrics lead to a one parameter family of connections and
curvature tensors such that Einstein's field equations are
modified as\f G_{\mu\nu}(E)=8\pi
G(E)T_{\mu\nu}(E)+g_{\mu\nu}(E)\Lambda (E).\label{MEE}\ff
Furthermore, in \cite{Magueijo02xx} the modified Schwarzschild
solution to (\ref{MEE}) has also been demonstrated in terms of
energy independent coordinates as \f ds^2=-{(1-{2GM\over r})\over
f_1^2}dt^2+{1\over f_2^2(1-{2GM\over r})}dr^2+{r^2\over
f_2^2}d\Omega^2 .\label{MBH}\ff From this solution we see that the
position of horizon  $R=2GM\sim M/ 4\pi M_p^2$ is fixed at the
usual place and universal for all observers, however in general
the area of the horizon  may be a function of
$f^2_2$ so as to be energy dependent.%In particular for
%this specific solution the area of the horizon is the same for all
%observers as well, although in general it may be a function of
%$f^2_2$ so as to be energy dependent.
\section{Thermodynamics of modified black holes}
Now we turn to investigate the thermodynamics of modified
Schwarzschild black holes. For explicitness, we adopt a modified
dispersion relation (MDR) by taking $f_1^2=[1-(l_pE)^2]$ and
$f_2^2=1$, where Planck length $l_p\equiv \sqrt{8\pi G}\equiv
1/M_p$. For this specific solution the area of the horizon is the
same for all observers as well. First we consider the temperature
of the modified black holes (\ref{MBH}), which  can usually be
obtained by calculating the surface gravity $\kappa$ on horizons,
namely \f T={\kappa \over 2\pi},\ff where $\kappa$ is related to
the metric by \f \kappa={-1\over 2}\lim_{r\rightarrow
R}\sqrt{-g^{11}\over g^{00}}{(g^{00})'\over g^{00}}.\ff

Now for modified Schwarzschild black holes it is straightforward
to obtain the surface gravity as \f \kappa={1\over
\sqrt{1-(l_pE)^2}}{1\over 4GM}.\ff Therefore, the temperature is
\f T={\kappa \over 2\pi}={1\over \sqrt{1-(l_pE)^2}}{1\over 8\pi
GM},\label{EDT}\ff which is also energy dependent.

Now we propose to define an intrinsic temperature for large
modified black holes by taking the photons in the vicinity of
black hole horizon as an ensemble. Suppose we make the measurement
with the use of photons with average energy $E=\langle E \rangle$,
then we expect that the temperature of black holes can be
identified with the energy of photons emitted from black holes
\cite{Adler01vs,Chen02tu}, namely $T=E$ such that we have \f
l_p^2T^4-T^2+{M_p^4\over M^2}=0.\ff

Thus we obtain the intrinsic temperature of modified Schwarzschild
black holes as %which has the form
\f T= \left[ {M_p^2\over
2}\left(1-\sqrt{1-{4M_p^2\over
M^2}}\right)\right]^{1/2}.\label{tem}\ff It requires that the mass
of black holes $M\geq 2M_p$, and correspondingly the temperature
$T\leq M_p/ \sqrt{2}$. For large black holes with $M\gg 2M_p$, it
goes back to the ordinary form, \f T= {M_p^2\over M}.\ff

The intrinsic temperature (\ref{tem}) is exactly the same as the
one we have obtained in \cite{LHL} through the modified dispersion
relation and uncertainty principle. The derivation there can be
summarized as follows.  Given $f_1^2=[1-(l_pE)^2]$ and $f_2^2=1$,
the corresponding modified dispersion relation reads as \f
l_p^2E^4-E^2+(p^2+m_0^2)=0,\label{n2}\ff thus the energy can be
non-perturbatively solved as \f E^2={1\over
2l_p^2}\left[1-\sqrt{1-4l_p^2(p^2+m_0^2)}\right]\label{mep}.\ff
Applying this relation to the photons emitted from black holes and
identifying the characteristic temperature of this black hole with
the photon energy $E$, we may have \f T= \left[ {M_p^2\over
2}\left(1-\sqrt{1-4l_p^2p^2}\right)\right]^{1/2}.\label{tep}\ff
Moreover, we apply the ordinary uncertainty relation to photons in
the vicinity of black hole horizons\cite{Adler01vs,Chen02tu}. \f
p\sim \delta p\sim {1\over\delta x}\sim {1\over 4\pi R},\ff where
a ``calibration factor'' $4\pi$ is introduced. Plugging this
relation into (\ref{tep}) we easily find the temperature of
Schwarzschild black holes is the same form as (\ref{tem}).

From (\ref{mep}) it is very interesting to notice that a single
particle has a maximum energy $E_{max}=M_p/\sqrt{2}$, rather than
$M_p$. This is important as it also provides a cutoff for the
temperature of modified black holes possibly probed by any
particle with energy $E$, which is (\ref{EDT}). From this equation
we notice that for a modified black hole with fixed mass M there
exists a maximum but finite value possibly probed any particle,
which is $T_{max}=\sqrt{2} M^2_p/ M$, rather than a divergent
number as one intuitively requires the energy of particles
approaches the Planck mass $M_p$.

Next assuming the first thermodynamical law holds for modified
black holes, namely $dM=TdS$, we may obtain the intrinsic entropy
by plugging the intrinsic temperature (\ref{tem}) into this
relation and then taking integrations. The result reads as
 \f S= {1\over \sqrt{2}}\left[ {A\over 8G}(1+t)^{3/2}-{1\over
\sqrt{2}}ln[{A\over 8G}(1+t)]\right],\ff where $t=\sqrt{1-8G/A}$.
When $A\gg 8G$, it becomes %the familiar formula
\f S={A\over 4G}-{1\over 2}ln{A\over 4G}+...\label{ec1}\ff
Therefore, we find the total entropy of modified black holes
contains a leading term which is nothing but the familiar
Bekenstein-Hawking entropy and a logarithmic correction term.

%This while but dispersion relations contributes a logarithmic
%correction to black hole entropy.

\section{Conclusions}
In this paper we have considered the thermodynamics of modified
black holes in the formulation of deformed general relativity.
Taking an specific rainbow metric as example, we find that the
temperature of black holes probed by particles is energy
dependent. But still we may define an intrinsic temperature as
well as entropy for such black holes by considering the ensemble
composed of photons in the vicinity of horizons. In contrast to
the ordinary ones, the mass of modified black holes is bounded
from below such that the temperature will reach a maximum but
finite value with the evaporation of black holes. In particular
they will cease radiation as the size of black holes approaches
the Planck scale, providing a mechanism to treat black hole
remnants as a candidate for cold dark
matter\cite{Barrow92hq,Adler01vs,Chen02tu,LHL}. The results
obtained in this paper are consistent with those obtained in
\cite{LHL} through the modified dispersion relations, and this
coincidence supports the deformed equivalence principle proposed
in \cite{Magueijo02xx}.

We point out that the scheme presented here can be generalized to
other sorts of modified black hole solutions and their
implications to cosmology are under investigation and will be
discussed elsewhere.

\section*{Acknowledgement}
 This work is partly
supported by NSFC (No.10405027, 10205002) and SRF for ROCS, SEM.
H.Zhang would like to thank the Center for Gravity and
Relativistic Astrophysics at Nanchang University for hospitality
during his recent visit.

\end{document}